\documentclass[12pt]{article}
\usepackage{amssymb,amsmath,slashed,cite}
\usepackage{color}

\makeatletter
\@addtoreset{equation}{section} %%% Latex Companion, p.~16
\makeatother

\newcommand{\e}{{e}}

\newcommand{\eqn}[1]{(\ref{#1})}
\def\appendix#1{\addtocounter{section}{1}\setcounter{equation}{0}
\renewcommand{\thesection}{\Alph{section}}
\section*{%Appendix~
\thesection\protect\indent \parbox[t]{11.715cm} {#1}}
\addcontentsline{toc}{section}{Appendix\thesection\ \ \ #1} }
\newcommand{\Tr}[1]{\:{\rm Tr}\,#1}
\def\one{\mbox{1 \kern-.59em {\rm l}}}
\newcommand{\be}{\begin{equation}}
\newcommand{\ee}{\end{equation}}
\newcommand{\beq}{\begin{equation}}
\newcommand{\eeq}{\end{equation}}
\newcommand{\bea}{\begin{eqnarray}}
\newcommand{\eea}{\end{eqnarray}}

\declareslashed{}{/}{.1}{0}{D}

\newcommand{\la}[1]{\label{#1}}
\newcommand{\ba}{\begin{eqnarray}}
\newcommand{\ea}{\end{eqnarray}}

\begin{document}
\begin{titlepage}
\begin{flushright}

\baselineskip=12pt
ICCUB-12-326\\
\hfill{ }\\
\end{flushright}

\begin{center}

\baselineskip=24pt

{\Large\bf Higgs-Dilaton Lagrangian from Spectral Regularization}

\baselineskip=14pt

\vspace{1cm}

M.A.~Kurkov$^{ab}$, Fedele Lizzi\,$^{abcd}$~\\
        
$^{a}$ Dipartimento di Scienze Fisiche, Universit\`{a} di
Napoli {\sl Federico II}~\\ $^b$ INFN, Sezione di Napoli\\
Monte S.~Angelo, Via Cintia, 80126 Napoli, Italy\\
 $^{c}$ High Energy Physics Group, Dept. Estructura i Constituents
de la Mat\`eria, \\Universitat de Barcelona, Diagonal 647, 08028
Barcelona, Catalonia, Spain \\
$^{d}$
Institut de Ci\`encies del Cosmos, UB, Barcelona\\
        {\small E-mail: {max.kurkov@na.infn.it,
        fedele.lizzi@na.infn.it}
        %Report \#
        }\\[10mm]

\end{center}

\vskip 2 cm

\begin{abstract}
In this letter we calculate the full Higgs-Dilaton action describing the Weyl anomaly 
using  the bosonic spectral action. This completes the work we started in our previous paper (JHEP 1110 (2011) 001). We also clarify some issues related to the dilaton and its role as collective modes of fermions under bosonization.
\end{abstract}

\end{titlepage}

\section{Introduction}
In~\cite{AndrianovKurkovLizzi1, AndrianovKurkovLizzi2} we discussed the idea of bosonization of all Standard Model fermion degrees of freedom into a single collective scalar field and we computed the effective Higgs dilation potential, expressing it via the 
bosonic spectral action \cite{SpectralAction}. The computation was performed
for constant fields and assuming a space-time to be flat and the dilaton to be spatially constant. This gives the main features of the potential, but does not describe completely the action since all derivative terms, including the kinetic term, were absent from the analysis. In this letter we present the complete calculation, to render the model implementable for investigation of the cosmic evolution. 

The presence of the dilaton field comes out from a spectral or finite mode regularization (FMR) and the anomaly bosonization. 
In \cite{AndrianovBonoraGamboa,AndrianovBonora1, AndrianovBonora2,AndrianovAxial} the mentioned approach was introduced and applied for the axial anomaly and in \cite{ Vassilevich} for the Weyl anomaly in case of massless fermions.\footnote{All these consideration however requires a compact spacetime, but the principle of a cutoff on the momentum eigenvalues is more general, see for example~\cite{Polchinski, poincareinvariant}.} The principle is based on the regularization of a partition function of a fermionic field theory whose fermionic action is generically written as
\be
S_F = \langle \Psi |D |\Psi \rangle = \int d^4 x \sqrt{g} \bar\psi(x) D_x \psi(x) \la{SF} 
\ee 
We will call the operator $D$ the generalized Dirac operator, or just the Dirac operator. It is a matricial operator which contains the usual Dirac operator, but also other terms, such as the mass of the particles. The role of this operator is crucial in the noncommutative geometry framework\cite{Connesbook}. The action~\eqn{SF} once inserted in a functional interaction integration diverges, and should be regularized, and the FMR, 
that counts only the eigenvalues of $D$ smaller than a cutoff scale $\Lambda$, spoils a Weyl invariance of the partition function. 

However in~\cite{AndrianovLizzi} it was shown that one can restore the Weyl invariance via
an addition to the classical action a slight variant of the spectral action which includes the dilaton. In contrast to the
approach \cite{AndrianovLizzi} and \cite{ChamseddineConnesscale}, in the present approach the dilaton is not considered as an independent field but is constructed from the Standard Model fermions via the anomalous bosonization.
Investigation on the consequences of the spectral action for cosmology~\cite{NelsonOchoaSakellariadou, BuckFairbairnSakellariadou, MarcolliPierpaoliTeh} are of great interest and this paper can be a starting contribution for the role of the dilaton. Moreover the applicability of the letter goes beyond the noncommutative geometry aspects, because the data we use is just is the action for fermions of the kind of the one for the standard model. 

The paper is organized as follows. In section~\ref{GENWinv} we show that the  Weyl transformed
generalized Dirac operator is nothing but the original operator computed with transformed vierbeins and Higgs fields.
In section~\ref{BOS} we recall the idea of the Weyl anomaly bosonization and
introduction of the dilaton field as a collective degree of freedom of all standard model fermions.
After that, using the observation from section~\ref{GENWinv} we finally compute all terms of the Higgs-dilaton potential in section~\ref{COMPUT}.
The last section contains conclusions. Several computational details are in two appendices.

\section{Generalized Weyl invariance \la{GENWinv}}
In~\cite{AndrianovKurkovLizzi2} we discussed the invariance properties of the fermionic action~\eqn{SF} under the following transformation:
\be
D\rightarrow \e^{-\frac{\phi}{2}}D \e^{-\frac{\phi}{2}},\quad \Psi \rightarrow \e^{\frac{1}{2}\phi}\psi,\quad 
\mbox{or} \quad D_x \rightarrow \e^{-\frac{5}{2}\phi}D_x \e^{+\frac{3}{2}\phi}, \quad
\psi  \rightarrow \e^{-\frac{3\phi}{2}}\psi,\la{TRANSzero}
\ee
where
\be
D\equiv g^{-\frac{1}{4}}D_x g^{\frac{1}{4}}\quad \Psi\equiv g^{-\frac{1}{4}}\psi \la{DvsDx}. 
\ee
The operator $D$ is the abstract operator on the Hilbert space of the vectors $\Psi$, while $D_x$ and $\psi(x)$ are their realizations as differential operators and functions on spacetime respectively.
The quantities $\psi$ and $D_x$ are usually used to write down the classical fermionic action with
a diffeomorfic invariant coordinate measure $d^4 x\sqrt{g}$, while the quantities $\Psi$ and $D$ are more 
convenient in the functional integral formalism, due to the diffeomorphic invariance of the measure $[d\Psi][d\bar\Psi]$ (see~\cite{Fujikawabook}).

The Dirac operator $D$ acts on left-right spinors as
\be
D_x=\left(\begin{array}{cc} D_G & \gamma_5 \otimes S\\
 \gamma_5 \otimes S^\dagger & D_G\end{array}\right) \label{Dstruct}
\ee
where $D_G$ is a "geometric" part of the Dirac operator,\footnote{Following a well established tradition, we use greek indexes to label coordinates, latin letters $k,l,m,n$ for Lorentz indexes and $a,b,c$ for gauge indexes.} 
\be
D_G = i \e^{\mu}_k \gamma^k \left(\partial_{\mu} - \frac{i}{2}\omega^{mn}_{\mu}\sigma_{mn} - i A_{\mu}^aT^a\right),
\ee
that contains the spin connection $\omega^{mn}_{\mu}$ and gauge fields $A_{\mu}$, 
and $S$ contains the
information about Higgs field $H$, Yukawa couplings, mixings i.e. all
terms which couple the left and right  spinors. The
gravitational background is in general nontrivial.  For the purpose of this note the relevant aspect is the presence on the Higgs field $H$, on which we concentrate our attention. For the following it is important to note that $S$ il \emph{linear} in $H$.

The important fact is that the transformation~\eqref{TRANSzero} is equivalent
to:
\be
g_{\mu\nu}\rightarrow \e^{2\phi}g_{\mu\nu},\quad \psi \rightarrow \e^{-\frac{3}{2}\phi}\psi,\quad H\rightarrow\e^{-\phi}H \la{TRANS}.
\ee 

The law of transformation of the Higgs field $H$ is in agreement with \eqref{TRANSzero}. 
To prove the equivalence of \eqref{TRANSzero} and \eqref{TRANS} finally,
we notice, that, under\footnote{More carefully one should write the corresponding transformation of vierbeins instead of the transformation of a metric tensor.} \eqref{TRANS}, the geometric part of the Dirac operator transforms as follows:
\be
D_G \rightarrow \e^{-\frac{5\phi(x)}{2}}D_G \e^{\frac{3\phi(x)}{2}}. \la{DGtrans} 
\ee
The mentioned result is present in~\cite{Fujikawabook}, however in the appendix A %\ref{appD} 
we give a more detailed proof.

We also remark that we \emph{do not} transform
the gauge fields $A_{\mu}^{a}$: they appear in $D_G$ multiplied by $\e^{\mu}_k$, so the correct transformation of the "gauge term" of $D_G$ is automatically provided by the transformation of the vierbeins.

\section{Generalized Weyl anomaly bozonization \la{BOS}}

Following~\cite{AndrianovKurkovLizzi2} we consider fermions, in a fixed gauge, Higgs and gravity background.
 Due to a generalized\footnote{Standard Weyl anomaly is related with the transformation of a metric tensor and fermions. Since now we also
transform the Higgs field $H$, now should  we call the anomaly``generalized". In the following for brevity we will skip the word "generalized". } Weyl anomaly, the fermonic partition function 
\be
Z_{F} = \int[d\Psi][d\bar{\Psi}] \e^{-S_F[\bar{\Psi},\Psi,\mbox{\scriptsize bosonic background}]}
\ee
is not invariant under \eqref{TRANS}. 
In~\cite{AndrianovKurkovLizzi2} it was shown that the Weyl non invariant part of $Z_F$ can be expressed in terms of the collective degree of freedom of all SM fermions,
the dilation:
\be
\int[d\Psi][d\bar{\Psi}] \e^{-S_F[\bar{\Psi},\Psi,\mbox{\scriptsize bosonic background}]} = \int[d\phi]  \e^{-S_{coll}[\phi,\mbox{\scriptsize bosonic background}] + W_{inv}}, \la{boso}
\ee
where $W_{inv}$ is (nonlocal) Weyl invariant functional of background fields, and $S_{coll}$ is a \emph{local} functional of background fields and
the dilation $\phi$. For a flat space-time and coordinate independent fields $S_{coll}$ was computed in~\cite{AndrianovKurkovLizzi2}. 

Let us  clarify some aspects of the introduction of the collective degree of freedom of all fermions,
or bosonization.
In our context the term ``bosonisation" does not mean that some composite operator $O_{\phi}(x)$, constructed from the 
 scalar field $\phi$ and its derivatives, equals another composite operator $O_{\Psi}(x)$, constructed from the fermionic fields $\Psi$ and $\bar\Psi$.  
More generally it means that the vacuum expectation of the product of $n$  bosonic composite operators $O_{\phi}(x)$
equals the vacuum expectation of the product of $n$
fermionic composite operators $O_{\Psi}(x)$ for $n=1,2,...$, i.e.\ equality of corresponding classes of Green functions.
\be \langle O_{\Psi}(x_{1}),...,O_{\Psi}(x_{n})\rangle_{\mathbf{ferm. ~vacuum}} = \langle O_{\phi}(x_{1}),...,O_{\phi}(x_{n})\rangle_{\mathbf{bos. ~vacuum}},\quad n=1,2,...\ee 

Now we will specify the mentioned classes of Green functions.  Substitute
$g_{\mu\nu} = \e^{2\alpha}g_{\mu\nu}$ and $H = \e^{-\alpha}H$ in \eqref{boso} and consider 
$\alpha$ as a source. Since the invariant part $W_{inv}$ in the right hand side of \eqref{boso}
remains unchanged under this substitution, it will not give contribution, one has:
\be
\left(\frac{\delta^n}{\delta \alpha(x_1)...\delta\alpha(x_n)}\log
Z_{F}^{\alpha}\right)
\Bigg|_{\alpha_1,...\alpha_n = 0} 
=\left( \frac{\delta^n}{\delta \alpha(x_1)...\delta\alpha(x_n)}\log Z_{coll}^{\alpha} \right)\Bigg|_{\alpha_1,...\alpha_n = 0},
\ee
where
\be
Z_F^{\alpha}\equiv\int[d\Psi][d\bar{\Psi}] \e^{-S_F[\bar{\Psi},\Psi,\e^{2\alpha}g_{\mu\nu},\e^{-\alpha}H]
},\quad Z_{coll}^{\alpha}\equiv\int[d\phi]  
\e^{-S_{coll}[\phi,\e^{2\alpha}g_{\mu\nu},\e^{-\alpha}H] 
}
\ee 
 
In our case the composite fermionic operator $O_{\Psi}$, that we bosonize, and the corresponding 
bosonic operator $O_{\phi}$ 
are given correspondingly by: 
\ba
O_{\Psi}(x)&=&\left(\frac{\delta}{\delta \alpha(x)} S_F[\bar{\Psi},\Psi,\e^{2\alpha}g_{\mu\nu},\e^{-\alpha}H]\right)_{\alpha=0},\la{Opsi}\\
O_{\phi}(x)&=&\left(\frac{\delta}{\delta \alpha(x)} S_{coll}[\phi,\e^{2\alpha}g_{\mu\nu},\e^{-\alpha}H]\right)_{\alpha=0}.
\ea

Notice that in the absence of the Higgs field, $H=0$, these operators
are nothing but (up to a $\sqrt{g}$ factor) traces of corresponding stress energy
tensors $T^{\mu\nu}_{F,coll}(x) = \frac{2\delta}{\sqrt{g}\delta g_{\mu\nu}(x)} S_{F,coll}$.
It is remarkable, that in this case the classical $T_{\mu~F}^{\mu}$ vanishes
on the equations of motion, however the quantum vacuum average $\langle T_{\mu~F}^{\mu}(x)\rangle_{\mathbf{ferm. vac.}}\neq 0$, due to the famous trace anomaly (see e.g. \cite{Fujikawabook}). The collective action 
describes the trace anomaly already on classical level:
\be
\langle T_{\mu~F}^{\mu}(x)\rangle_{\mathbf{ferm. vac.}} = \langle  T_{\mu~coll}^{\mu}(x)\rangle_{\mathbf{bos. vac.}} \simeq T_{\mu~coll}^{\mu}(x)\big|_{\phi = \phi_{class}} + \mbox{loop corrections}, 
\ee
where $\phi_{class}(x)$ solves the classical equations of motion $\frac{\delta S_{coll} [\phi]}{\delta{\phi(x)}} = 0$.
In contrast  to the fermionic partition function, the bosonic partition function doen't poses the trace anomaly,
and the Weyl non invariance of action appears already at classical level. 

In the presence of the Higgs field, i.e.\ when the Dirac operator is given by \eqref{Dstruct}, 
the operator $O_{\Psi}(x)$, given by \eqref{Opsi} equals to
\be
O_{\Psi} = \sqrt{g}\left(T_{\mu~F}^{\mu}  - \gamma_5\otimes S(H)\bar\psi\psi\right),\quad T_{\mu~F}^{\mu}\equiv \frac{2g_{\mu\nu}}{\sqrt{g}}\frac{\delta}{\delta g_{\mu\nu}} S_F,
\ee
 besides $\langle T_{\mu}^{\mu}\rangle$  now $\langle O_{\Psi}\rangle$ contains an additional
 fermionic condensate $\langle \bar\psi(x)\psi(x)\rangle$ contribution.

In the next section we will evaluate the collective action $S_{coll}$ expressing it through the (modified) bosonic spectral action.
This computation is strongly based on
the use of the heat-kernel expansion, that being an asymptotic expansion, strictly makes sense in the weak fields approximation, and faces problems beyond it \cite{IochumLevyVassilevich}. Nevertheless the bosonization that we discuss is also valid in low energy region,
that justifies the use of the heat kernel in our treatment.

\section{Computation of Higgs-dilaton Action. \la{COMPUT}}
Following our previous work, and taking into account that now the dilaton field depends on space-time coordinates, we have:
\be
S_{coll} = -\left(1-\Lambda^2\log\frac{\Lambda^2}{\mu^2}\partial_{\Lambda^2}\right)
\int^{1}_0 dt \Tr\left\{\phi\chi\left( \frac{(\e^{-\frac{\phi t}{2}} D\e^{-\frac{\phi t}{2}})^2}{\Lambda^2}\right)\right\} \la{Scoll2}
\ee
Since we perform the finite mode regularization, that cuts all eigenvalues of $D$ higher than the cutoff scale $\Lambda$, we actually deal
with a \emph{bounded} operator, therefore under the sign of $\Tr$ 
we can replace $\left(\tilde{D}\right)_{\phi t}\equiv \e^{-\frac{\phi t}{2}} D\e^{-\frac{\phi t}{2}}$ by $\left(\tilde{D}_x\right)_{\phi t}\equiv \e^{-\frac{5}{2}\phi t}D_x \e^{+\frac{3}{2}\phi t}$,  
where $D_x\equiv g^{-\frac{1}{4}}D g^{\frac{1}{4}} $ (c.f.\eqref{SF}).
As shown in  Sec. \ref{GENWinv} 
we can rewrite \eqref{Scoll2} in the following way:
\be
S_{coll} = -\left(1-\Lambda^2\log\frac{\Lambda^2}{\mu^2}\partial_{\Lambda^2}\right)
\int^{1}_0 dt \left(\Tr\left\{\phi\chi\left(  \frac{D_x^2}{\Lambda^2}\right)\right\}\right)\bigg|_{g_{\mu\nu} = \left(\tilde{g}_{\mu\nu}\right)_{\phi t},~H = \left(\tilde H\right)_{\phi t}} \la{Scoll}
\ee
Using the heat kernel expansion one can show, that (for details see for example~\cite{SpectralAction})
\ba
\Tr\left\{\phi\chi\left(  \frac{D_x^2}{\Lambda^2}\right)\right\} &=& \int d^4 x \sqrt{g} \phi \left(\frac{45\Lambda^4}{8\pi^2} + \frac{15\Lambda^2}{16\pi^2}\left(R - 2y^2 H^2\right) \right.\nonumber\\
&&+\frac{1}{4\pi^2}\left(\frac{3}{8}R_{;\mu}^{~~\mu} +\frac{11}{32}G_B- y^2 H_{;\mu}^{2~\mu}+3y^2\left(D_{\mu}HD^{\mu}H - \frac{1}{6}R H^2\right)\right.
  \nonumber \\
&& \left.\left.
+3z^2H^4 + G_{\mu\nu}^iG^{\mu\nu i} 
+W_{\mu\nu}^{\alpha}W^{\mu\nu\alpha} 
+\frac{5}{3}B_{\mu\nu}B^{\mu\nu} - \frac{9}{16}C_{\mu\nu\rho\lambda}C^{\mu\nu\rho\lambda}\right)\right)\nonumber\\ \la{SBphi} ,
\ea
where $y^2$ and $z^2$ stand for correspondingly quadratic and quartic combinations of the
Yukawa couplings, whose precise definition can be found for example inö\cite[Eq.~3.17]{SpectralAction}. Since the Yukawa couplings are strongly dominated by the one of the top quark $Y_t$, one can keep in mind, that $y^2 \simeq Y_t^2, z^2\simeq Y_t^4$. 
$G_B$ denotes the Gauss-Bonnet density:
\be
G_B\equiv \frac{1}{4}\epsilon^{\mu\nu\rho\sigma}\epsilon_{\alpha\beta\gamma\delta}R^{\alpha\beta}_{\mu\nu}
R^{\gamma\delta}_{\rho\sigma}.
\ee
Substituting the expression \eqref{SBphi} into \eqref{Scoll} we finally get the result:
\ba
S_{coll}\equiv \int d^4 x \sqrt{g}\left(A\left(\e^{4\phi}-1\right) + B H^2\left(\e^{2\phi}-1\right) - C\phi H^4 
-\alpha_1\left(\e^{2\phi}-1\right)R + \alpha_2 \e^{2\phi}\left(\phi_{;\mu}\phi_{;}^{~\mu}\right) \right. \nonumber \\
- \alpha_3~\phi\left(3y^2\left(D_{\mu}H D^{\mu}H - \frac{1}{6}R H^2\right) 
+  G_{\mu\nu}^i G^{\mu\nu i} + W_{\mu\nu}^{\alpha}W^{\mu\nu\alpha} + \frac{5}{3}B_{\mu\nu}B^{\mu\nu} - \frac{9}{16}C_{\mu\nu\rho\lambda}C^{\mu\nu\rho\lambda}\right)\nonumber \\
\left.-\alpha_4\left(12R\left(\phi_{;\mu}^{~\mu}+\phi_{;\mu}\phi_{;}^{~\mu}\right)+11\phi G_B + 44G^{\mu\nu}\phi_{;\mu}\phi_{;\nu}
+14\left(\phi_{;\mu}^{~\mu}+\phi_{;\mu}\phi_{;}^{~\mu}\right)^2 + 22\left(\phi_{;\mu}^{~\mu}\right)^2\right)\right)\nonumber\\
\la{finalansw},
\ea 
where $G_{\mu\nu}$ stands for the Einstein tensor and the constants $A$, $B$, $C$, $\alpha_1..\alpha_4$, are defined as follows: 
\ba
A &=& \left(2\log{\frac{\Lambda^2}{\mu^2}} - 1\right)\frac{45 \Lambda^4}{32\pi^2}, \quad
B = \left(1 - \log{\frac{\Lambda^2}{\mu^2}}\right)\frac{15 \Lambda^2 y^2}{16\pi^2},\quad
C = \frac{ 3z^2}{4\pi^2}, \\
\alpha_1 &=&\left(1 - \log{\frac{\Lambda^2}{\mu^2}}\right)\frac{15\Lambda^2}{32\pi^2},\quad
\alpha_2 =  \left(1 - \log{\frac{\Lambda^2}{\mu^2}}\right)\frac{45\Lambda^2}{16\pi^2}, \quad
\alpha_3 = \frac{1}{4 \pi^2}, \quad
\alpha_4 = \frac{1}{128\pi^2  }.\nonumber
\ea
For technical details of this calculation see the appendix% \ref{app2}.

\section{Conclusions}
The full Higgs-dilaton Lagrangian is computed, in the general case, thus completing the program me started in~\cite{AndrianovKurkovLizzi1, AndrianovKurkovLizzi2},
the result given by the expression~\eqref{finalansw}. We opens the gate for a complete analysis of the cosmological consequences of the model.  We may already say that for a choice of the normalization point parameter $\mu$, that
provides existence of minimum of the Higgs-dilaton potential, kinetic term of the dilaton field comes out with the correct positive sign. In the local minimum point $\phi = \phi_{m}$ the induced $R$ term appears also with the correct positive plus.
We have also clarified the role of the bosonic fields vs.\ the fermionic ones, and given a prescription to recognize the bosonization.  

%IThe $O(\Lambda^0)$ term contains negative defined contributions, that are however 
%of order $k^2/\Lambda^2$, where $k$ is a typical momentum of the field $\phi$ and $\Lambda\sim\Lambda_{Pl}$.  
%Since the bosonization is considered as a low energy effect with respect to $\Lambda_{Pl}$, these terms can be neglected in the region of 
% applicability.
%There exists a natural choice of the normalization point $\mu = \e^{-\frac{1}{4}}\Lambda$, that removes a huge $\sim \Lambda^4$ contribution
%to the cosmological constant. 

\setcounter{section}{0}
\appendix{ Equivalence of Transformations  \la{appD}}

In this appendix we give the details of the equivalence of the two transformations~\eqn{TRANSzero} and~\eqn{TRANS}.

The Weyl transformation of the metric tensor in terms of vierbeins is given by:
\be
e_{\mu k} \rightarrow \e^{\phi(x)} e_{\mu k}, ~~\e^{\mu k} \rightarrow \e^{-\phi(x)} e^{\mu k}.\la{Vtrans}
\ee 
The spin connection $\omega^{mn}_{\mu}$ has the following expression via vierbeins (see \cite{Fujikawabook}):
\be
\omega^{mn}_{\mu} = \frac{1}{2}e^{m\lambda}e^{n\rho}\left(C_{\lambda\rho\mu} - C_{\rho\lambda\mu} - C_{\mu\lambda\rho}\right),\la{spin}
\ee
where 
\be
C_{\lambda\rho\mu} = e^k_{\lambda}(\partial_{\rho}e_{k\mu} - \partial_{\mu}e_{k\rho}). \la{C}
\ee
Substituting the transformation \eqref{Vtrans} in \eqref{C} and \eqref{spin} we find the following law of the spin connection transformation
under \eqref{Vtrans}:
\be
\omega_{\mu}^{mn}\rightarrow \omega_{\mu}^{mn} + e^{m}_{\mu}e^{n\rho}\partial_{\rho}\phi - e^{n}_{\mu} e^{m\lambda}\partial_{\lambda}\phi.
\ee
Now we remind, that the generators of the representation of Lorentz group $\sigma_{mn}$ for spin $1/2$ have the following form
in terms of the Dirac matrixes:
\be
\sigma_{mn} = \frac{i}{4}\left[\gamma_m,\gamma_n\right].
\ee

Therefore one has the following law of transformation the combination $\frac{i}{2}\omega^{mn}_{\mu}\sigma_{mn}$ (this formula is presented in~\cite{Fujikawabook}\footnote{Comparing our and~\cite{Fujikawabook} formulas: note that Fujikawaand and Suzuki use $\alpha(x)$ = -$\phi(x)$}):
\be
\frac{i}{2}\omega^{mn}_{\mu}\sigma_{mn} \rightarrow \frac{i}{2}\omega^{mn}_{\mu}\sigma_{mn} - \frac{1}{2}\gamma_{\mu}\gamma^{\alpha}\partial_{\alpha}\phi
+\frac{1}{2}\partial_{\mu}\phi.
\ee
So we have:
\be
-\gamma^{\mu}\frac{i}{2}\omega^{mn}_{\mu}\sigma_{mn}\rightarrow -\gamma^{mu}\frac{i}{2}\omega^{mn}_{\mu}\sigma_{mn} + \frac{3}{2}\gamma^{\mu}\partial_{\mu}\phi. \la{termtrans}
\ee
Finally, using \eqref{termtrans} and \eqref{Vtrans} we obtain:
\ba
D_G &\equiv&i e^{\mu}_k \gamma^k \left(\partial_{\mu} - \frac{i}{2}\omega^{mn}_{\mu}\sigma_{mn} - i A_{\mu}^aT^a\right)\rightarrow \nonumber\\
&\rightarrow& i e^{\mu}_k \gamma^k  e^{-\phi}  \left(\partial_{\mu} - \frac{i}{2}\omega^{mn}_{\mu}\sigma_{mn} - i A_{\mu}^aT^a +\frac{3}{2}\partial_{\mu}\phi\right) = \nonumber \\
&=& i e^{\mu}_k \gamma^k e^{-\frac{5\phi}{2}}\left(\partial_{\mu} - \frac{i}{2}\omega^{mn}_{\mu}\sigma_{mn} - i A_{\mu}^aT^a\right)e^{+\frac{3\phi}{2}}
 = e^{-\frac{5}{2}\phi(x)}D_{G}e^{+\frac{3}{2}\phi(x)}.
\ea

\appendix{Computational details.\la{app2}}
\subsection{$R$ contribution.}
Under the Weyl transformation of the metric tensor,
\be g_{\mu\nu}\rightarrow \left(g_{\mu\nu}\right)_{\phi}\equiv e^{2\phi}g_{\mu\nu} \la{Wtrans}
\ee 
the scalar curvature transforms as follows:
\be
R\rightarrow \left(\tilde R\right)_{\phi}\equiv e^{-2\phi}\left(R + 6\left( \phi^{~\mu}_{;\mu}  + \phi_{;\mu}\phi_{;}^{~\mu}\right) \right),\la{Rtrans}
\ee
and integrating by parts one can easily show, that
\be
-\int d^4 x ~\phi \int_0^1dt \left(\sqrt{\tilde g} \tilde R\right)_{\phi t} 
=\int d^4 x\sqrt{g}\left(-\frac{1}{2} \left(e^{2\phi} - 1\right)R  + 3\cdot e^{2\phi}\left(\phi_{;\mu}\phi_{;}^{~\mu}\right)\right).
\ee

\subsection{$\left(H^2\right)_{;\mu}^{~\mu}$ and $R_{;\mu}^{~\mu}$ contributions. }
For a scalar quantity $f$, that 
transforms under the Weyl transformation \eqref{Wtrans} as 
\be
f\rightarrow \left(\tilde f\right)_{\phi}, 
\ee
its Laplacian $\Delta f \equiv \nabla_{\mu}\nabla^{\mu} f$ transforms as follows:
\be
\Delta f \rightarrow \left(\tilde\Delta\tilde f\right)_{\phi} 
= e^{-4\phi}\nabla^{\mu}e^{2\phi}\nabla_{\mu} \left(\tilde f\right)_{\phi }. \la{Lapltrans} 
\ee

For $f = H^2$ and $f = R$, using the transformation law for $H$ that can be inferred from~\eqn{TRANSzero}, and relations \eqref{Wtrans} and \eqref{Lapltrans}, we obtain the following contributions to the Higgs-dilaton potential:
\be
-\int {d^4 x}~\phi \int_0^1dt \sqrt{\tilde g} \left(\tilde\Delta\tilde H^2\right)_{\phi t}
= -\int d^4 x \sqrt{g}\left( \phi^{~\mu}_{;\mu}  + \phi_{;\mu}\phi_{;}^{~\mu}\right)H^2
\ee
and
\be
-\int d^4 x~\phi \int_0^1dt \left( \sqrt{\tilde g}\tilde\Delta \tilde R\right)_{\phi t } 
=- \int d^4 x \sqrt{g}\left(\left( \phi^{~\mu}_{;\mu}  + \phi_{;\mu}\phi_{;}^{~\mu}\right)R 
+ 3\left(\phi_{;\mu}^{~~\mu} + \phi_{;\mu}\phi_{;}^{~\mu}\right)^2\right).
\ee 

 \subsection{$G_B$ contribution.}
It is known from the differential geometry, that the Gauss-Bonnet density $G_B$ in a four dimensional space-time can be presented in the following form, convenient for the forthcoming analysis: 
\be
G_B =  C_{\mu\nu\rho\sigma}C^{\mu\nu\rho\sigma} 
- 2\left(R_{\mu\nu}R^{\mu\nu} - \frac{1}{3}R^2\right) \la{GBsqr}.
\ee 

 For the transformed Ricci tensor under the Weyl transformation \eqref{Wtrans} we have:
 \be
 \left(\tilde{R}_{\mu\nu}\right)_{\phi } = R_{\mu\nu} + 2\left(\phi_{;\mu\nu}- \phi_{;\mu}\phi_{;\nu}\right)  + \left(\phi_{;\lambda}^{~ \lambda} +
 2\phi_{;\lambda}\phi_{;}^{~\lambda}\right)g_{\mu\nu}  \la{Riccitrans} 
 \ee 
Using laws of transformations of the Ricci tensor and the scalar curvature 
\eqref{Riccitrans}, \eqref{Rtrans} and also Weyl invariance of the Weyl tensor contribution after some simple computations we obtain:
 \be
\sqrt{g} G_B \rightarrow \left(\sqrt{\tilde{g}} \tilde{R}^*\tilde{R}^*\right)_{\phi} = \sqrt{g}\left(G_B + \nabla_{\mu} J^{\mu}\right).
\ee 
where the current $J^{\mu}$ is defined as follows:
 \be
 J^{\mu} \equiv 8\left(-\phi_{\nu}G^{\nu\mu} + \left(\phi_{;\lambda}^{~\lambda} + \phi_{;\lambda}\phi_{;}^{~\lambda}\right)\phi_{;}^{~\mu}\right) 
 - 4\left(\phi_{;\lambda}\phi_{;}^{~\lambda}\right)_{;}^{~\mu},\la{currdef}
 \ee

Contribution of the Gauss-Bonnet term to the Higgs-dilaton potential is propotional (with the sign plus) to the following expression:
 \ba
 -\int{d^4x} \phi(x)\int_0^1{dt}\left(\sqrt{\tilde{g}}\tilde{R}^*\tilde{R}^* \right)_{\phi t} = \nonumber\\
 \int{d^4x\sqrt{g}}\left(-\phi G_B- 4 G^{\mu\nu}\phi_{;\mu}\phi_{;\nu} + 2\left(\phi_{;\mu}\phi^{~\mu}_{;}\right)^2 
 + 4\left(\phi_{;\mu}\phi^{~\mu}_{;}\right)\phi_{;\lambda}^{~\lambda}\right).
 \ea
 
\paragraph{Acknowledgments}
We are grateful to A.A.~Andrianov for discussions and suggestions. FL acknowledges
support by CUR Generalitat de Catalunya under project FPA2010-20807 and the {\sl Faro} project {\sl Algebre di Hopf, differenziali e di
vertice in geometria, topologia e teorie di campo classiche e
quantistiche} of the Universit\`a di Napoli {\sl Federico II}.

\end{document}